\documentstyle[preprint,aps]{revtex}

\newcommand{\referencestyle}{
\small
\abovedisplayskip=6pt
\belowdisplayskip=6pt
\vspace{12pt}}

\def\D{\Delta}

\def\Dt{\Delta\tau}

\def\t0{\tau_0}

\def\t{{\tau}}

\def\ben{\begin{eqnarray}}
\def\enn{\end{eqnarray}}
\def\ov{\over\displaystyle\strut}
\def\dst{\displaystyle\strut}
\def\l({\left(}
\def\r){\right)}
\def\o{{out}}
\def\s{{side}}

\begin{document}
\include{epsf}
\begin{center}
{ \Large\bf
Bose-Einstein Correlations for Expanding Finite Systems \\}
or\\
{\large \bf  From a Hot Fireball to a Snow-Flurry}
\end{center}
\medskip
\begin{center}
        B. L\"orstad\footnote{Presentation of works done together with  T.
Cs\"org\H o}
\end{center}
\medskip
\begin{center}
{\it
Department of Physics \\
University of Lund,
S\"olvegatan 14,
S - 223 62 Lund, Sweden \\
e-mail:bengt.lorstad@quark.lu.s
}
\end{center}
%\vfill
\begin{abstract}
Most boson emitting sources contain a core of finite dimensions
surrounded by a large halo, due to long-lived resonances like
$\omega,\eta,\eta',K^{0}$ etc. When the Bose-Einstein correlation (BEC)
function
of the core can be determined we show that its intercept ($\lambda$) measures,
as a function of momentum, the square of the fraction of core
particles produced. A simultaneos measurement of BEC and
the single-particle distributions can thus determine the characteristics
of the core.

There are two types
of scales present in the space-like as well as in the
time-like directions in a model-class describing a cylindrically
symmetric, finite, three-dimensionally expanding boson source.
One type of the scales is related to the finite lifetime and
geometrical size of the system, the other type is goverened
by the change of the local momentum distribution in temporal and spatial
direction.
If the geometrical sizes of the core are sufficiently large
the parameters of the BEC function obey
the $m_{t}$-scaling observed in $SPb$ and $PbPb$ reactions at CERN.
The model can describe the measurements of the single- and two-particle
distributions in the central region of $SPb$ reactions. A fit to experimental
data shows
that the freeze-out of hadrons occurs at a larger volume and at a
much lower temperature than that given by the measurement of the inverse slope
of the $m_{t}$-spectrum and standard BEC analysis.

\end{abstract}

\section{Introduction.} Hanbury-Twiss correlations were discovered 40 years
ago, revealing information about the angular diameters of distant
stars~\cite{HBT}. The method, also referred to as intensity interferometry,
extracts information from the quantum statistical correlation function
for (partially) chaotic fields. The method has extensively been applied
to the study of the freeze-out geometry in high energy nucleus-nucleus
collisions as well as in elementary particle reactions. For recent reviews
see refs.~\cite{bengt,zajc}.

Accumulating evidence indicates that the space-time structure of the pion
emission
in heavy ion reactions at the 200 AGeV bombarding energy region at CERN SPS
has a peculiar feature, namely that the boson emission can be approximately
 divided into two parts: the centre and the
halo~\cite{halo1,halo2,halo3,halo4}.
The central part corresponds to a direct production mechanism
e.g. hydrodynamical evolution or particle production from excited strings,
followed by subsequent rescattering of the particles, while
the surrounding halo corresponds to pions emitted from the decay of
long-lived hadronic resonances, like $\omega, \eta$, $\eta'$ and $K^{0}$,
which have a mean decay length of more than 20 fm.
These long-lived resonances will decay and give rise to mainly low momentum
pions.
However, the momentum distribution of the emitted pions is independent
from the position of the decay which in turn is smeared out over a large
region due to the long mean decay length.
This mechanism will result in pions of similar (low) momentum from
a very large volume. This is in contrast to the core which when
expanding in a hydrodynamical manner will have pions of similar momentum
from a relatively small volume.
We will investigate the phenomenological consequences
of such a structure where there is a characteristic length-scale
of the core which is smaller than any length-scale that can be connected
to the surrounding halo. In the case of halo length-scales larger than 20 fm
 they give rise to a sharp peak of the correlation function in the $Q \le
\hbar/R \approx 10$
MeV/c region. This region is strongly influenced by the Coulomb interaction so
the
accuracy of the correlation function determination will be dependent on proper
Coulomb corrections which constitute a problem in itself. Furthermore, this
$Q < 10$ MeV/c region is very difficult to measure due to that we must properly
determine two very close tracks which make experimental systematic errors
largest
in this region of the correlation function even for high resolution experiments
like
the NA35 and NA44 high energy heavy ion experiments at CERN.

We will show  that even if we cannot measure reliably the correlation function
below
$Q<10$ MeV/c important new information about the particle production in a
core/halo
scenario can be given by analyzing both the invariant momentum distribution of
single particles
and the correlation function for $Q>10$ MeV/c. We will apply our approach to
NA44 data on
$SPb$ reactions at 200 AGeV/c.

We investigate the core/halo scenario
 utilizing analytical results in the Wigner-function formalism.  This
formalism is
well suited to describe both the invariant momentum distribution (IMD) and
the Bose-Einstein correlation function (BECF). We utilize the version of the
formalism
discussed in refs.~\cite{pratt_csorgo,zajc} and applied to analytic
calculations
in refs.~\cite{lutp,nr,1d,3d,uli_s,uli_l} recently.

\section{The Wigner-function formalism}
In the Wigner-function formalism,
the one-boson emission is characterized by the emission function, $S(x,p)$,
which is the time derivative of the single-particle Wigner function. Here
$x = (t,\vec r\,) = (t, r_x, r_y, r_z)$ denotes the four-vector in space-time,
and $p = (E, \vec p \,) = (E, p_x, p_y, p_z)$ stands for the four-momentum
of the on-shell particles with mass $m = \sqrt{ E^2 - \vec p^{\, 2} }$.
The transverse mass is defined as $m_t = \sqrt{m^2 + p_x^2 + p_y^2}$
and $y = 0.5 \log\l({\dst E+p_z \ov E-p_z} \r) $ stands for the rapidity.
Attention is drawn to the point that the normal size character $x$ stands for a
four-vector,
normalsize $y$ is the symbol for the rapidity
while the subscripts $_x $ and $_y$ index directions in coordinate and momentum
space.

For chaotic sources the two-particle Wigner-function can be expressed
in terms of the symmetrized products of the single-particle Wigner functions.
Based on this property, the two-particle Bose-Einstein correlation functions
can be determined from the single-particle  emission function solely.
These relations are especially simple in terms of the Fourier-transformed
emission function,
\ben
\tilde S(\Delta k , K ) & = & \int d^4 x \,\,
                 S(x,K) \, \exp(i \Delta k \cdot x ),\label{e:ax}
\enn
where
\begin{eqnarray}
\Delta k  = p_1 - p_2, & \quad \mbox{\rm \phantom{and}}\quad &
K  = {\displaystyle\strut p_1 + p_2 \ov 2}
\end{eqnarray}
and $\Delta k \cdot x $ stands for the inner product
of the four-vectors.
The  invariant momentum distribution of the emitted particles
is given by
\ben
 {\dst d^2 n \ov dy \, dm_t^2 } & = & \tilde S(\Delta k = 0, K =
p).\label{e:imd}
\enn
The IMD is normalized to unity as
\ben
\int dy \, dm_t^2 {\dst d^2 n \ov dy \, dm_t^2 } \,\, & = & 1.\label{e:nimd}
\enn
The two-particle
 BECF-s are prescribed in terms of our  auxiliary
function,
\ben
C(K,\Delta k) & \simeq & 1 +  {\displaystyle\strut
         \mid \tilde S(\Delta k , K) \mid^2 \ov \mid \tilde S(0,K)\mid^2 },
\enn
 as was presented e.g. in ref. \cite{pratt_csorgo,zajc}.
The effect of final state  Coulomb and Yukawa
interactions
shall be neglected as implicitly assumed by the above
equation,
and we assume completely chaotic  emission.

The Wigner-functions are in general complex valued functions, being the
quantum analogue of the classical phase-space distribution functions.
  For the purpose of Monte-Carlo simulations of the Bose-Einstein correlation
functions the off-shell Wigner-functions are approximated by the on-shell
classical emission functions from  the simulation~\cite{pratt_csorgo,RQMD};
 in case of analytic
calculations they are usually modeled by some positive valued
functions~\cite{lutp,nr,1d,3d,uli_s,uli_l}.
Neither of these approximations seems to be necessary:
in fact we need only simplifying assumptions as listed below.

\section{The core/halo model}
{\it Assumption 0}. There is a resolvable length-scale in the core

{\it Assumption 1}. The bosons are emitted either from a
central {\it core} or
from the surrounding  {\it halo}. The respective emission functions
are indicated by $S_c(x,p)$ and $S_h(x,p)$, and $f_c $ indicates the
 fraction of the bosons emitted from the central part.
According to this assumption, the complete emission function can be written as
\ben
 S(x,p) = f_c  S_c(x,p) + ( 1 - f_c) S_h(x,p).
\enn
Both the emission function of the core and that of the halo are normalized
similarly to the complete emission function, normalized by
eqs.~(\ref{e:ax},\ref{e:imd},\ref{e:nimd}).

{\it Assumption 2}. We assume that the emission function which
characterizes
the halo changes on a scale $R_H$ which is larger
 than  $R_{max}\approx \hbar / Q_{min}$, the maximum length-scale resolvable
by the intensity interferometry microscope. $Q_{min}$ is taken as 10 MeV/c in
our
considerations. However, the smaller central part
of size $R_c$ is
assumed to be resolvable,
\ben
 R_H > R_{max} > R_c.
\enn
This inequality is assumed to be satisfied by all characteristic scales
in the halo and in the central part, e.g. in case the side, out or longitudinal
components~\cite{bertsch,lutp} of the correlation function are not identical.

Based on {\it Assumption  1} the IMD can be re-written as
\ben
{\dst d^2 n \ov dy \, dm_t^2 } = f_c \, {\dst d^2 n_c \ov dy \, dm_t^2 } +
                                ( 1 - f_c) \, {\dst d^2 n_h \ov dy \, dm_t^2 },
        \label{e:ich}
\enn
where the subscripts $c,h$ index the contributions by the central and the
halo parts  to the IMD given  respectively by
\ben
{\dst d^2 n_i \ov dy \, dm_t^2 } & = & \tilde S_i(\Delta k = 0, K = p)
\enn
for $i = c,h$.
The normalization condition according to {\it Assumption 1} reads as
\ben
\int dy \, dm_t^2 {\dst d^2 n_i \ov dy \, dm_t^2 } \,\, & = & 1\label{e:nimdi}
\enn
for $i = c,h$.
Note that the IMD as expressed by eq.~(\ref{e:ich}) includes the possibility
that the IMD for the bosons of the halo is different from the bosons emerging
from the central part. Thus the relative contribution of the halo and the
core is a function of the momentum in this model.

The BECF is expressed by
\ben
C(K,\Delta k) = 1 +
 {\dst \mid f_c \, \tilde S_c(\D k, K) + ( 1-f_c) \, \tilde S_h(\D k, K) \mid^2
\ov
       \mid f_c \, \tilde S_c(\D k=0, K=p) + ( 1-f_c) \, \tilde S_h(\D k=0,
K=p) \mid^2 },
        \label{e:cmat}
\enn
which includes interference terms for boson pairs of $(c,c)$ $(c,h)$ and
$(h,h)$ type.
Due to the assumption that the emission is completely chaotic, the exact value
of the BECF at the $\D k = 0$ is always 2 in this model.

The measured two-particle BECF is determined for
$Q>Q_{min}\approx 10$ MeV/c,
and any structure within the $ \D k < Q_{min}$ region cannot be resolved.
However, the $(c,h)$ and $(h,h)$ type boson pairs create a narrow peak
in the BECF exactly in this $\D k $ region according to eq.~(\ref{e:cmat}),
which cannot
be resolved due to {\it Assumption 2}. According to this assumption,
 the Fourier-transformed emission function of the halo for non-zero
relative momenta vanishes at the given resolution $Q_{min}$.
At zero relative momentum the same quantity gives the single particle
momentum
distribution from the halo, which is not effected by the {\it two}-particle
resolution.

Including the finite resolution effect, symbolized by the horizontal bar,
 the measured BECF can be written as
\ben
\overline{ C(\D k, K) } & = & 1 +
      \lambda_{*}(K=p;Q_{min}) {\dst \mid \tilde S_c( \D k, K) \mid^2 \ov
                             \mid \tilde S_c( \D k = 0, K=p) \mid^2},
                           \label{e:lamq}
\enn
where
\ben
\lambda_{*}(K=p;Q_{min})= {\dst \mid f_c \tilde S_c( \D k=0, K=p) \mid^2 \ov
                                   \mid f_c \, \tilde S_c(\D k=0, K=p) + (
1-f_c) \, \tilde S_h(\D k=0,K=p)\mid^2},
\enn
or expressed with the IMDs
\ben
\lambda_{*}(p;Q_{min}) = {\dst \mid f_c \, {\dst d^2 n_c \ov dy \, dm_t^2 }
\mid^2 \ov
                        \mid  \, {\dst d^2 n \ov dy \, dm_t^2 } \mid^2}.
\enn
The IMD is connected to the cross-sections by the relations
\ben
 {\dst 1 \ov \sigma}  {\dst d^2 \sigma_{i} \ov dy \, dm_t^2 }=
 <n_{i}> {\dst d^2 n_{i} \ov dy \, dm_t^2 }
\enn
for $ i=c,h$. From Eq. (8) we deduce $f_{c}={\dst <n_{c}> \ov <n>}$. The
effective intercept
$\lambda_{*}(p;Q_{min})$ is then also given by

\ben
\lambda_{*}(p;Q_{min}) = {\dst \mid \, {\dst d^2 \sigma_c \ov dy \, dm_t^2 }
\mid^2  \ov
                        \mid  \, {\dst d^2 \sigma \ov dy \, dm_t^2 } \mid^2}.
\enn

The effective intercept parameter $\lambda_{*}$ shall in general  depend on the
mean momentum of the observed boson pair, and we have $0 \le \lambda_{*} \le
1$.
The equation above shows that this {\it effective $\lambda_{*}$ has a very
simple interpretation
as the square of the fraction of core particles to all particles emitted}.
We also see that this equation makes it possible to measure the cross-section
of the bosons produced
in the central core with the help of the combined use of the BECF parameter
$\lambda_{*}(y,m_{t})$
and the cross-section of all particles.

Note that the $\lambda_{*}(y, m_t)$ intercept parameter in our
picture reflects the drop in the BECF
from its exact value of 2 due to the contribution from the
resonance halo, when measured with the finite two-particle momentum
resolution
$Q_{min}$. Thus the measured intercept,
$\lambda_{*}(y, m_t)$ does {\it not} coincide with the
exact value of the BECF at zero relative momentum
within this description.

The effect of the halo is to introduce a momentum-dependent effective intercept
parameter to the
BECF. The shape of the BECF for $Q>Q_{min}$ shall be solely determined by the
freeze-out
phase-space distribution in the central part. This central part or core is
usually well accessible
to hydrodynamical calculations.

Note also that this result
 has in principle
nothing to do with the similar relation obtained for the
case of partially coherent, partially chaotic fields, where $\lambda_{*} =
f_{inc}^2$,
and $f_{inc} = <n_{inc} > / < n_{tot}> $ gives the fraction of the total
multiplicity
in the chaotic field, although the relation formally is similar.
For the case of partially coherent fields, the Bose-Einstein
correlation function contains an interference term in between the coherent and
the chaotic
fields which leads to a double-Gaussian or double-exponential
structure~\cite{gyu_ka},
emphasized recently in ref.~\cite{weiner}.
In our case, the BECF contains not three but only two terms,
and it looks very much like the BECF would look
like for the case the halo were missing i.e. $f_c = 1$.

Note also, that the focus is {\it not} on the reduction of the intercept of the
BECF
due to the halo (resonance decays) but more precisely on the {\it
momentum-dependence}
of this reduction. The fact of the reduction has been known before~
\cite{halo2,halo1,pratt_csorgo,RQMD}, however its momentum dependence was not
utilized as far as we know. Within our formalism however it is straightforward
to show that the core fraction of bosons can be estimated by
\ben
f_c & = & \int_{-\infty}^{\infty} dy \int_{m^2}^{\infty} dm_t^2
\sqrt{\lambda(y,m_t) }
                {\dst d^2 n \ov dy \, dm_t^2 }  \label{e:d}
\enn
and we also have
\ben
\min \sqrt{\lambda(y,m_t)}\, \le \, f_c \, \le  \, \max \sqrt{\lambda(y,m_t)}.
\enn

\underline{\it Application}.
Present NA44 data~\cite{na44mt} for central $S + Pb$ reactions at CERN SPS
with
200 AGeV bombarding energy show an approximately $m_t$ independent  intercept
parameter:
$\lambda_{\pi^+} = 0.56 \pm 0.02$ and $0.55 \pm 0.02$ at the quite different
mean
transverse momenta of $150 $ MeV and $ 450 $ MeV, respectively. This suggests
that
in the midrapidity region the momentum distribution of the resonance halo is
similar to that of the central part, and the halo contains $ 1 - \sqrt{\lambda}
=
25 \pm 2 $ \% of all the pions.
For this special case of $\lambda = f_c^2 = const$ we have the following
simplified equations for the IMD and the BECF:
\ben
{\dst d^2 n \ov dy \, dm_t^2 } \, & = & \, {\dst d^2 n_c \ov dy \, dm_t^2 }
                                \, = \, {\dst d^2 n_h \ov dy \, dm_t^2 },\\
\overline{ C(\D k, K) } \,  & = & \, 1 + f_c^2 R_c(\D k, K).
\enn
I.e. the only apparent effect of the halo is to reduce the intercept
parameter of the measured BECF to $\lambda  = f_c^2$ while the IMD and
the $(\D k, K)$ dependence of the BECF is determined by the central
part {\it exclusively}.

Pions coming from the resonance-halo surrounding a strongly
interacting centre are predominant at lower values of the transverse momentum
according to the SPACER calculations in ref.~\cite{lutp} and RQMD calculations
in ref.~\cite{RQMD}, similarly to
the results of calculations with HYLANDER~\cite{halo2} and the hydrodynamical
calculations of the Regensburg group~\cite{uli_h}. The resonance fraction
as a function of the transverse momentum was explicitly shown in the
publications
{}~\cite{RQMD,halo2}. Each of the mentioned models predicts that the halo of
resonances
produces predominantly low momentum pions, and they predict that the effective
 intercept $\lambda_{*}(p;Q_{min})$ should increase with increasing
transverse mass in contrast to the measured values.

The constancy of the effective parameter $\lambda_{*}(y,m_{t})$ with respect to
$m_{t}$ suggests a mechanism of enhancing the low momentum particles in the
core.
Such an enhancement has a natural explanation in a hydrodynamical description
of
an expanding core as we will present below, see also  \cite{qm95,3d}.
 The mechanism is a simple volume
effect, as pointed out in \cite{uli_l}.
 To calculate the cross-section
you integrate over the volume from where the particles of a given momentum
are emitted. This volume is measured by the BECF. In the NA44 experiment
 all three radii of this volume are equal and they are
inversely proportional the the square root of $m_{t}$~\cite{na44mt}. This will
give
an extra volume factor, $V_{*}\propto (m_{t})^{-3/2}$, enhancing the single
particle
cross-section at low $m_{t}$. We will return to this point.

\underline{\it Discussion.} The general result for the BECF of systems with
large halo
 coincides with the most frequently applied phenomenological parameterizations
 of the BECF in high energy heavy ion as well as
in high energy particle reactions~\cite{bengt}.
Previously, this form has received a lot of criticism from the theoretical
side,
claiming that it is in disagreement with quantum statistics~\cite{weiner}
or that the $\lambda $ parameter is just a kind of fudge parameter
or a measure of our ignorance,
which has been introduced to make theoretical predictions comparable to data.
Now one can see that this type of parameterizations can be derived
with a standard inclusion of quantum statistical effects -- if we assume that
we discuss interferometry for systems with large halo. The fact that most of
the
studied reactions, including $e^+ + e^-$ annihilations,
various lepton-hadron and hadron-hadron reactions, nucleon-nucleus and
nucleus - nucleus collisions are phenomenologically well
describable~\cite{bengt}
with variations of eq.~(12) thus does not exclude the possibility
that these high energy reactions all create boson emitting systems which
include a large halo.

The applicability of the halo picture to
a given reaction is not necessarily the only one possible explanation
of a reduced intercept. It is known that the final state interactions
may have an influence on the effective intercept of the two-particle
correlation functions~\cite{bowler} and the effect has been shown to
result in stronger drop in the intercept value for smaller source
i.e. for hadronic strings created in lepton-lepton, lepton-hadron
or hadron-hadron collisions. However, more detailed calculations
indicate that the higher order corrections for the final state interactions
may very well cancel the first order effects resulting in a very small
total final state interaction correction~\cite{bow}.
In our calculations these effects have been neglected, so our results
should in principle be compared to data which present the {\it genuine}
Bose-Einstein correlation function. This function is not necessarily the
same as the short-range part of the two-particle correlation function
since Coulomb and Yukawa interactions as well as the $\eta' \eta$ decay chain
and other short-range correlations may mask the quantum statistical
correlation functions. However, as the particle density is increased
(in the limit of very large energy or large colliding nuclei),
the Bose-Einstein correlations together with the final state interactions
 dominate over other short-range correlations
due to combinatoric reasons.

\section{the model of the core}
For central heavy ion collisions at high energies the beam or $z$ axis
becomes a symmetry axis. Since the initial state of the reaction is
axially symmetric and the equations of motion do not break this pattern,
the final state must be axially symmetric too.
However, in order to generate the thermal length-scales in  the transverse
directions, the flow-field must be either three-dimensional,
or the temperature must have significant gradients in the
transverse directions. Furthermore, the local temperature
may change during the the duration of the particle
emission either because of the re-heating of the system caused by the
hadronization and/or intensive rescattering processes
or the local temperature may decrease because of the expansion
and the emission of the most energetic particles from the
interaction region.

We model the emission function of the core for high energy heavy ion reactions
as
as
\begin{eqnarray}
S_c(x,K) \, d^4 x & = & {\dst  g \ov (2 \pi)^3} \,  m_t \cosh(\eta - y) \,
\exp\l( - {\dst K \cdot u(x) \ov  T(x)} + {\dst \mu(x) \ov  T(x)}\r)
  \, H(\tau) d\tau  \, \tau_0 d\eta \, dr_x \, dr_y, \\
u(x) & \simeq & \l( \cosh(\eta)
        \l(1 + b^2 \, {\dst r_x^2 + r_y^2 \ov 2 \tau_0^2}\r) ,
        \,\, b \,{\dst r_x \ov \tau_0}, \,\, b \,{\dst r_y \ov  \tau_0}, \,\,
        \sinh(\eta)
        \l( 1 + b^2 \,{\dst r_x^2 + r_y^2 \ov 2 \tau_0^2} \r) \, \r),\\
T(x) & = & {\dst T_0
	\ov \l( 1 + a^2 \, {\dst  r_x^2 + r_y^2 \ov 2 \tau_0^2} \r)
	   \, \l( 1 + d^2 \, {\dst (\tau - \tau_0)^2 \ov 2 \tau_0^2  } \r)} \\
{\dst \mu(x) \ov T(x) }  & = &  {\dst \mu_0 \ov T_0} -
        { \dst r_x^2 + r_y^2 \ov 2 R_G^2}
        -{ \dst (\eta - y_0)^2 \ov 2 \Delta \eta^2 }.
\end{eqnarray}
We include a finite duration,  $H(\tau) \propto \exp(-(\tau-\tau_0)^2
/(2 \Delta\tau^2))$,        $\tau =\sqrt{t^2 - z^2}$, $\tau_0$ is the mean
emission time,
        $\Delta \tau$ is the duration of the emission in (proper) time,
        $u(x)$ is the four-velocity of the expanding matter,
        $\mu(x)$ is the chemical potential and $T(x)$ is the local
        temperature characterizing the particle emission.

	This emission function corresponds to
	a Boltzmann approximation to the local momentum distribution
	of a longitudinally expanding finite system which expands
	into the transverse directions with a non-relativistic
	transverse flow.
The decrease of the temperature in the transverse direction is
controlled by the parameter $a$, while
the strength of the transverse flow is controlled by the parameter $b$.
	The parameter $c = 1$ is reserved to denote the speed of light,
	and the parameter $d$ controls the strength of the change of the
	local temperature during the course of particle emission.

	For the case of $a = b = d = 0$ we recover the case of longitudinally
	expanding finite systems. The finite geometrical and temporal
	length-scales are represented by the transverse geometrical size
	$R_G$, the geometrical width of the space-time rapidity distribution
	$\Delta \eta$ and by the mean duration of the particle
	emission $\Delta \tau$. We assume here that the finite geometrical
	and temporal scales as well as the transverse radius and proper-time
	dependence of the inverse of the local temperature can be
	represented by the mean and the variance of the respective variables
	i.e. we apply a Gaussian approximation, corresponding to the
	forms listed above, in order to get analytically trackable results.
	We have first proposed the $a = 0, b = 1$ and $d = 0$ version
 	of the present model, and elaborated also the $a = b = d =0$ model
	~\cite{1d} corresponding to  longitudinally expanding finite
	systems with a costant freeze-out temperature and no transverse flow.
	Soon the parameter $b$ has been introduced~\cite{uli} and it has
	been realized that the transverse flow has to be non-relativistic
	at the saddle-piont corresponding to the maximum of the emission
	function. Yu. Sinyukov and collaborators classified the various
	classes of the ultra-relativistic transverse flows~\cite{hhm:te},
	~\cite{akkelin},  and introduced a parameter which controls
	the transverse temperature profile, corresponding to the
	$a \ne b = 0$ case. We have studied~\cite{qm95} the
	model-class $ a, b, d=0$ which we extend here to the $d \ne 0$
	case too.

 The integrals of the emission function  are evaluated
using the saddle-point method ~\cite{hhm:te,akkelin,uli}.
The saddle-point equations are solved in the LCMS~\cite{1d}, the longitudinally
comoving system,  for $\eta_s <<1 $ and $r_{x,s} << \tau_0 $.
These approximations are warranted if
$ |y - y_0| << 1 + \Delta \eta^2 m_t /T_0$ and
 $\beta_t
 = p_t / m_t << (a^2 + b^2) / b$.
The flow is non-relativistic at the saddle-point if $\beta_t
  << (a^2 + b^2) / b^2$.
The radius parameters are evaluated here up to
${\cal O}(r_{x,s}/\tau_0) + {\cal O}(\eta_s)$, keeping only the leading order
terms in the LCMS. However, when evaluating the invariant momentum
distribution (IMD),
sub-leading terms coming from the $\cosh(\eta-y)$ pre-factor
are also summed up,
since this factor influences the IMD in the lower $m_t$ region where
 data are accurate.

\subsection{Bose-Einstein correlations }
The experimental BECF  is parameterized in the form of
$C(Q_L,Q_{side},Q_{out}) = 1 + \lambda \exp(-R_L^2 Q_L^2 - R_{side}^2
Q_{side}^2 - R_{out}^2 Q_{out}^2) $ where the intercept parameter
$\lambda$ and the radius parameters may depend on the rapidity
and the transverse mass of the pair.
	The parameters of the the correlation function
	are given by
\ben
	R_{\s}^2 & = & R_*^2, \label{e:side} \\
	R_{\o}^2 & = & R_*^2 + \beta_T^2 \Dt_*^2 ,\label{e:out} \\
	R_L^2 & = & \tau_0^2 \Delta\eta_*^2 \label{e:long}
\enn
We obtain~\cite{1d,3d}
\ben
{\dst 1 \ov R_*^2 } & = &
		{\dst 1 \ov R_G^2} +
		{\dst 1 \ov R_T^2 } \cosh[\eta_s]  \\
{\dst 1 \ov \Delta \eta_*^2} & = &  {\dst 1 \ov \Delta \eta^2 } +
                {\dst 1 \ov \Delta \eta_T^2} \cosh[\eta_s] -
			{\dst 1 \ov \cosh^2[\eta_s]}, \\
{\dst 1 \ov \Delta \tau_*^2 } & = &
		{\dst 1 \ov \Delta\tau^2} + {\dst 1 \ov \Delta\tau_T^2}
			\cosh^2[\eta_s].
\enn
where the thermal length-scales are given by
\ben
 R_T^2 & = & { \displaystyle\strut \tau_0^2 \over
	\displaystyle\strut a^2 + b^2 }
	 { \displaystyle\strut T_0 \over \displaystyle\strut m_t}, \\
 \Delta\eta_T^2 & = & {\dst T_0 \ov m_t},\\
 \Delta\tau_T^2 & = & {\dst \tau_0^2 \ov d^2} {\dst T_0 \ov m_t}.
\enn
As a consequance, the parameters of the BECF-s are dominated
by the smaller of the geometrical and the thermal scales
not only in the spatial but in the temporal directions too.
These analytic expressions indicate that the BECF views only
a part of the space-time volume of the expanding systems,
which implies that even a complete measurement of the
parameters of the BECF as a function of the mean momentum $K$
may not be sufficient to determine uniquely
 the underlying phase-space distribution.
We also can see that  for pairs with $\mid y_0 - Y \mid << 1 + \Delta\eta^2
M_t/T_0 $
 the terms arising from the non-vanishing values of $\eta_s$ can be neglected.
In this approximation, the cross-term generating hyperbolic mixing angle
 $\eta_s \approx 0$
thus we find the leading order LCMS result:
\ben
	C(Q;K) & = & 1 + \lambda_*(K) \exp( - R_L^2 Q_L^2 - R_{\s}^2 Q_{\s}^2
			  - R_{\o}^2 Q_{\o}^2) ,
\enn
	with a vanishing out-long cross-term, $R_{\o,L} = 0$. $\lambda_*(K)$ is given
by eqs. (14) or (16). According to the previous part, $\lambda_*(K)=f_c^2 $ in
case
of the NA44 measurements.

	The difference of the side and the out radius
	parameters is dominated by the lifetime-parameter $\Delta\tau_*$.
	Thus vanishing difference in between the $R_\o^2 $ and $R_\s^2$
	can be generated dynamically in the case when the duration
	of the particle emission is large, but the thermal duration
	$\Delta\tau_T$ becomes sufficiently small. This in turn can be associated
	with intensive changes in the local temperature distribution
	during the course of the particle emission.
	If the finite source sizes are large compared to the thermal
	length-scales we have
\ben
	\Dt_*^2 & = & \Dt_T^2 \, {\dst 1 \ov 1 + {\dst \Dt_T^2 \ov \Dt^2} }
		\approx {\dst \tau_0^2 \ov d^2} \, {\dst T_0\ov m_t} \\
	R_L^2 & \approx  & {\tau_0^2 } \, {\dst T_0\ov m_t} \\
	R_\s^2 & \approx  & {\dst \tau_0^2 \ov a^2 + b^2 } \, {\dst T_0\ov m_t}
\enn
	Thus if $d^2 >> a^2 + b^2 \approx 1$ the model features
	a dynamically generated vanishing difference between
	the side and out radii.

	If the vanishing duration parameter is generated dynamically,
	 the model predicts an $m_t$ - scaling for the life-time parameter
	as
\ben
	\Delta\t_*^2 & = & {\dst R_\o^2 - R_\s^2 \ov \beta_t^2 } \simeq {\dst 1 \ov
			m_t},
\enn
	Note that this prediction could be checked experimentally
	if the  error bars of the
        measured radius parameters could be decreased to such a level that
	the difference between the out and the side radius parameters
	could be significant.

	If the finite source sizes are large compared to the thermal
length-scales and if we also have $a^2 + b^2 \approx 1$,
one obtains an
$m_t$ -{\it scaling} for the parameters of the BECF,
\ben
R_{side}^2 &\simeq & R_{out}^2 \simeq R_L^2 \simeq \tau_0^2 {\dst T_0 \ov m_t},
\quad \mbox{\rm valid for} \quad \beta_t << {\dst (a^2 + b^2) \ov b^2}
\simeq {1 \ov b^2}.
\enn

	Note that this relation is independent of the particle type and has been
seen in the recent NA44 data~\cite{na44mt}.
This $m_t$-scaling may be valid to arbitrarily large
transverse masses with $\beta_t \approx 1$ if $b^2 << 1$.
Thus, to generate the vanishing difference between the
side and out radius and the $m_t$ scaling simultaneously,
the parameters have to satisfy
	$ b^2 << a^2 + b^2 \approx 1 << d^2$,
i.e. the fastest process is the cooling, the next dominant
	process within this phenomenological picture has to be the development
	of the transverse temperature profile and
	finally the transverse flow shall be relatively weak.

\subsection{Invariant momentum distributions}
The IMD plays a {\it complementary role}
to the measured Bose-Einstein correlation function ~\cite{nr,1d,3d}.
Namely, the width of the rapidity distribution at a given $m_t$
as well as
$ T_*$ the effective temperature at a mid-rapidity $y_0$ shall be
dominated by the {\it longer} of the thermal and geometrical
length-scales.
Thus a {\it simultaneous analysis} of the Bose-Einstein correlation function
and the IMD may reveal information
both on  the temperature and flow profiles  and on the geometrical sizes.
E.g. the following relations hold:
\begin{eqnarray}
\Delta y(m_t)^2 & = & \Delta \eta^2 + \Delta \eta_T^2(m_t), \qquad
\mbox{\rm and} \qquad
{\dst 1 \ov T_*}  =  {\dst f \ov T_0 + T_G(m_t=m) } + {\dst 1 -f \ov T_0}.
\end{eqnarray}
The geometrical contribution to the effective temperature is given by
$T_G  = T \,  R_G^2 / R_T^2$ and the fraction $f$ is defined as
$f = b^2 / (a^2 + b^2)$, satisfying $0 \le f \le 1$.
The saddle-point sits at $\eta_s =
(y_0 - y) / (1 + \Delta\eta^2 / \Delta \eta_T^2 )$, $r_{x,s} =
\beta_t b R_*^2 / (\tau_0 \Delta\eta_T^2)$,  $r_{y,s} = 0$ and $\tau_s = \tau_0
$.

For the considered model, the invariant momentum distribution of the core
can be calculated as
\ben
{\dst d^2 n_c \ov dy \, dm_t^2 } & = &
	{\dst g \ov (2 \pi)^3 } \,\, \exp\l( {\dst \mu_0 \ov T} \r) \, \,
	m_t \,\, (2 \pi \Delta\eta_*^2 \tau_0^2)^{1/2} \, \,
	(2 \pi R_*^2) {\dst \Delta \tau_* \ov \Delta \tau} \,\, \cosh(\eta_s) \, \,
	\exp(+ \Delta \eta_*^2 / 2) \times  \nonumber \\
\null & \null & \times \exp\l( - {\dst (y - y_0)^2 \ov 2 (\Delta\eta^2 +
	\Delta \eta_T^2) } \r) \,
\exp\l( - {\dst m_t \ov T_0} \l(  1 - f \, {\dst \beta_t^2 \ov 2} \r) \r) \,
\exp\l( - f \, {\dst m_t \beta_t^2 \ov 2 (T_0+ T_G)} \r).
\label{e:imd2}
\enn
This  IMD has a rich structure:
it features { both a rapidity-independent
and  a rapidity-dependent low-pt enhancement}
as well as a
{ high-pt enhancement or decrease}.

\section{Applications to NA44 data}

We have fitted the NA44 preliminary
invariant momentum distributions for pions and kaons together
with the final NA44 data~\cite{na44mt} for the $m_t$ dependence of the
BECF parameters for both pions and kaons. As $\lambda$ has been shown not to
be dependent on $m_{t}$ the IMDs measured are proportional to the IMDs of the
core.
Fixing the parameter $y_0 = 3$
we get a description of the IMD and the BECF for both pions and kaons.
The result of the preliminary analysis, see Table, indicates
large geometrical source sizes for the pions, $R_G(\pi) \approx 7$ fm and $R_L
\approx 20$ fm,
late freeze-out times of $\tau_0 \approx 7$ fm/c, a vanishing duration
$\Delta \tau$ for both pions and kaons and finally surprisingly
low freeze-out temperatures, $T_0 \approx 100 $ MeV.
The kaons appear from a smaller region of $R_G(K) \approx 4$ fm.

 Let us mention,
that the measured points for the IMD contain statistical errors only,
systematic
errors are not included. This implies that the errors of the parameters, as
stated
in the Table, are probably to be enlarged, when we can use final data.
The $\Delta\chi^2$ is relatively large from the pion invariant momentum
distribution. We have scanned the parts which contribute to the large
increase in $\chi^{2}$ and concluded that this increase seems to be related to
some
large fluctuations of data points at the edge of the acceptance region.

\section{Conclusions}
 In summary we have studied the case when the central boson-emitting
region
is surrounded by a large halo, which also emits bosons. If the size of the halo
is
so large that it cannot be resolved in Bose-Einstein correlation measurements,
 lot of information shall be concentrated in the momentum dependence of the
intercept parameter of the correlation function.
We have shown that with the help of the Bose-Einstein correlation
measurement, the invariant momentum distribution can be measured for the
two independent components
belonging to the core and the halo, respectively.
The results do not depend on any particular parametrization of the core
nor of the halo.

Analysis of the NA44 data for two-pion correlations
indicated that the normalized invariant momentum distribution
of the pions from the halo of long-lived resonances
within errors coincides with the normalized invariant momentum
distribution of the pions from the central core.
This result can be explained with a hydrodynamical model of the core
development
in space-time.
The number of pions coming from the halo region, $\ge 20$ fm,
 was found to be $ 25 \pm 2$ \%
of the total number of pions within the NA44 acceptance.

 Instead of observing a small fireball
we are observing a big and  expanding snow-flurry
at CERN SPS $S + Pb$ reactions accordingly to this
analysis of
the partly preliminary NA44 data.
The central temperature in  the snow-flurry is slightly higher
than in the outer regions, however the temperature gradient is rather
small ($a^2 = 0.03 \pm 0.01$). The temporal changes of the local
temperature during the particle emission seem to be rather intensive ($ d^2 =
0.9 \pm 0.3$).

{\it Acknowledgments:} I want to express my sincere gratitude to the organizers
of this
Symposium for creating such a nice and stimulating atmosphere.

\newpage
\begin{center}
Table\\
Results of a simultaneous fit to NA44 preliminary
invariant momentum distributions and Bose-Einstein radius parameters
for $\pi$ and $K$ in central $S + Pb$ reactions at 200 AGeV

\vspace{5mm}
\begin{tabular}{|c|c|}  \hline
$\chi^{2}/ndf$, full fit & 490/220 \\  \hline
$\Delta \chi^{2}$, pion singles(128) &351 \\
$\Delta \chi^{2}$, kaon singles(94) &137 \\
$\Delta \chi^{2}$, radii measurements(9) &  2 \\ \hline
$T_{0}$ [MeV] &  100$\pm$2 \\
$\tau_{0}$ [fm/c]  &  6.9$\pm$0.2 \\
$R_{G(\pi)}$ [fm]  & 6.9$\pm$0.4 \\
$R_{G,(K)}$ [fm]  & 4.1$\pm$0.4 \\
$R_{G,L}$ [fm]  &  21$\pm$6 \\
$\Delta \tau(\pi)$ [fm/c] &0.1$\pm$0.1 \\
$\Delta \tau(K)$ [fm/c]  &  6$\pm$4 \\
$a^{2}$ &0.03$\pm$0.01 \\
$b^{2}$ & 0.85$\pm$0.01 \\
$d^{2}$ & 0.9$\pm$0.3 \\ \hline
\end{tabular}
\vspace{5mm}
\end{center}

\vfill
\eject
\end{document}